\documentclass[conference]{IEEEtran}
\usepackage[pdftex]{graphicx}
\begin{document}

\title{CoAP over ICN}

\author{\IEEEauthorblockN{Nikos Fotiou$^1$, Hasan Islam$^2$, Dmitrij Lagutin$^2$, Teemu Hakala$^3$, George C. Polyzos$^1$}
\IEEEauthorblockA{$^1$Athens University of Economics and Business, $^2$Aalto University, $^3$ELL-i} 
}

\maketitle

\begin{abstract}
The Constrained Application Protocol (CoAP) is a specialized Web transfer protocol for resource-oriented applications intended to run on constrained devices, typically part of the Internet of Things. In this paper we leverage Information-Centric Networking (ICN), deployed within the domain of a network provider that interconnects, in addition to other terminals, CoAP endpoints in order to provide enhanced CoAP services. We present various CoAP-specific communication scenarios and discuss how ICN can provide benefits to both network providers and CoAP applications, even though the latter are not aware of the existence of ICN. In particular, the use of ICN results in smaller state management complexity at CoAP endpoints, simpler implementation at CoAP endpoints, and less communication overhead in the network.     
\end{abstract}

\IEEEpeerreviewmaketitle

\section{Introduction}
The Internet of Things (IoT) is expected to interconnect billions of (usually constrained) devices that will generate vast amounts of information. Significant efforts have been devoted into enabling smart devices to connect to the Internet, share information, and consume services. These efforts have resulted in a variety of network access technologies and higher layer protocols. On the other hand, core (inter-)networking technologies have not been adapted to this new paradigm, raising concerns about whether or not networks will be able to cope with the scale and the patterns of the traffic of the IoT. In order to assuage these concerns a number of researches have sprung up proposing Future Internet (FI) architectures. One such promising FI architecture is Information-Centric Networking (ICN). \footnote{A survey of ICN research and architectures that have been investigated, with some still being pursued and experimentally explored further, can be found in~\cite{Xyl2013}.} ICN advocates implementing all (inter-)networking functions around content (i.e., information) identifiers, rather than location identifies. This shift in focus and techniques is expected to overcome various limitations of the current Internet~\cite{Tro2010}. However, such a shift requires not only the re-design of networking protocols, but also the modification of legacy Internet applications. Such radical changes at all network layers are an overwhelming barrier to the adoption of ICN. With this in mind, the POINT project~\cite{Tro2015} proposes a radical approach to ICN adoption: it postulates an individual ICN operator that uses network attachment points that translate \emph{legacy} IP applications traffic to ICN, i.e., the endpoints are oblivious to ICN.


In this paper, we explore how the Constrained Application Protocol (CoAP) can benefit from the POINT architecture, as well as, how a network operator that offers CoAP connectivity can benefit from ICN. By presenting some intriguing CoAP communication patterns, we show that information-centrism facilitates CoAP services and components development. Moreover, we show that by leveraging the multicast nature of POINT an operator can achieve significant benefits in terms of communication overhead. 

The reminder of this paper is organized as follows. In Section~\ref{sec:back} we describe the CoAP protocol, as well as, the information-centric POINT architecture. In Section~\ref{sec:coap} we illustrate CoAP over ICN reference architecture, alongside with the implementation details. Section~\ref{sec:coap} also highlights how ICN can benefit CoAP-based applications. In Section~\ref{sec:plat} we present the CoAP-specific extensions to the POINT platform. Finally,  Section~\ref{sec:conc} concludes our paper.

\section{Background}
\label{sec:back}
\subsection{CoAP}
CoAP~\cite{rfc7252} has been designed and developed to be a 'lightweight HTTP' so that it can be suitable to operate in constrained IP networks. The CoAP interaction model is similar to the client/server model of HTTP: a CoAP client issues a request message to a server and if the CoAP server is able to serve the request, it responds to the requester with a response code and the payload. Unlike HTTP, CoAP requests and responses are exchanged asynchronously, on top of an unreliable datagram oriented transport protocol (e.g., UDP). A CoAP message may carry a Token which is used to correlate an asynchronous response with the corresponding request. 
 
The CoAP protocol also supports intermediaries and caching of responses. There are two different kinds of proxies: Forward-Proxy and Reverse-Proxy. A Forward-Proxy sends a request to the server on behalf of a client. For this, the Forward-Proxy needs to be configured to perform requests on behalf of the client. In contrast, a Reverse-Proxy is transparent to the client. The Reverse-Proxy behaves as if it were the origin server. CoAP includes support for the discovery of resources exploiting a separate entity called Resource Directory (RD) which stores the descriptions of resources. Moreover, CoAP supports group communication~\cite{rfc7390} based on IP multicast; CoAP groups and the membership of a group can be discovered via the lookup interfaces in the Resource Directory (RD). Finally, CoAP enables clients to observe a resource through a simple publish/subscribe mechanism ~\cite{rfc7641}. With this, the server asynchronously pushes the notification of state changes of the resource for which the client is interested in and follows a best-effort approach to gurantee the eventual consistency of the observed state and the actual state of the resource. 
 
\begin{figure}
\includegraphics[width=0.95\linewidth]{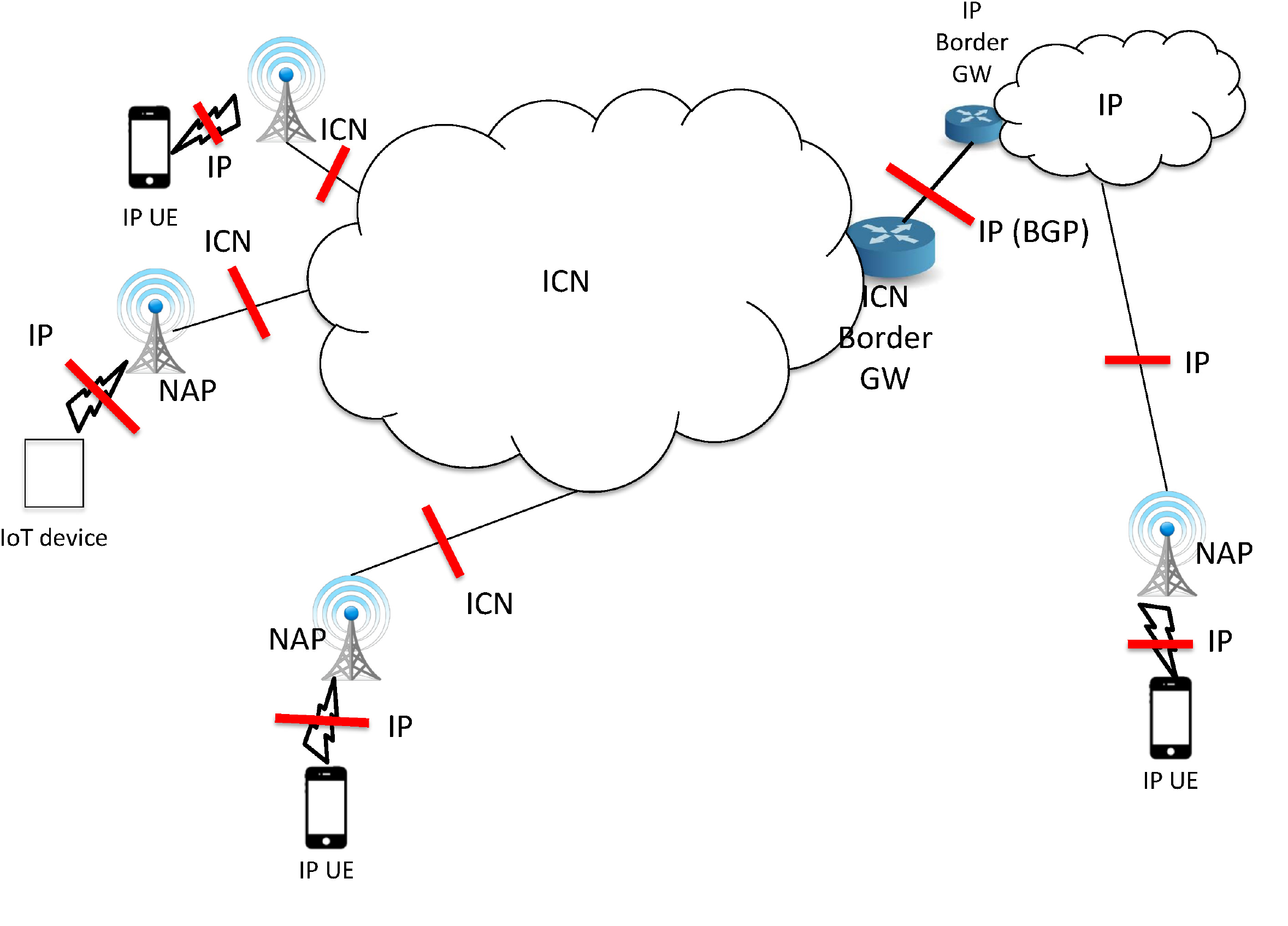}
\caption {The POINT architecture.}
\label{fig:point}
\end{figure}

\subsection{The POINT architecture}

The POINT architecture interconnects IP endpoints using the Publish-Subscribe Internet (PSI) ICN architecture~\cite{Xyl2012}. 
The PSI architecture is based on the publish-subscribe 
paradigm  where  users  interested  in  receiving  some  content 
subscribe  for it through their network device, referred to as the 
subscriber, and content owners store their content on a network 
device which advertises it and if requested publishes
it (hence these devices are referred to as the 
publishers). Every content item is identified by a flat identifier known 
as  the Rendezvous  Identifier (RId).  Moreover,  every  content 
item belongs to at least one scope. The purpose of a scope is to 
give a hint about content location and to group content items 
with  the  same  dissemination  level.  Scopes  are  hierarchically
organized and identified by a Scope Identifier
(SId). Scopes are managed by specialized Rendezvous Nodes (RNs), which form an  overlay 
Rendezvous  Network.  The  rendezvous  network provides a lookup service, which routes a subscription 
to a RN that ``knows'' (at least) one publisher for the requested item.
A typical  transaction  in  PSI  involves  the  following
steps. An owner of a content item assigns it
a RId and stores a copy of it in at least one publisher that 
advertises its availability  in  one  or  more  scopes. With  this 
advertisement, the RId is stored in the RNs that manage these 
scopes.  Subscribers  send  subscriptions  for  specific   (SId,RId) 
pairs, which are routed by the rendezvous network 
to an appropriate RN. Upon receiving a subscription
message and provided that at least one publisher exists, the RN instructs 
a Topology  Manager to  create  a  forwarding  path  from  a 
publisher to the subscriber, which is included in the notification 
message to the publisher. Finally, the 
content item is transferred from the publisher to the subscriber.     

Instead of dictating a clean-slate end-to-end ICN architecture, which would be very challenging to deploy, POINT allows standard IP traffic to be run over an ICN core network in a more efficient way~\cite{Tro2015}. To achieve this, the POINT architecture (Figure~\ref{fig:point}) provides a number of handlers for existing IP-based protocols (e.g., HTTP, CoAP, basic IP) that map the underlying protocols onto appropriate named objects within the ICN core. Therefore existing applications can benefit from ICN's features such as native multicase and caching. The potential benefits of the POINT architecture compared to IP-based ones are highlighted in more detail in~\cite{Tro2015}. 

Standard end-user devices are connected to the POINT architecture through network attachment points (NAPs) where protocols handlers 
are implemented. Inside the operator's core network, the named objects generated by the protocol handlers are advertised, published and 
forwarded using standard ICN functions. When necessary, a named object is translated into the appropriate IP/HTTP/CoAP packet
and it is transmitted back to the end-user device.

%

\section{CoAP over ICN}
\label{sec:coap}
In this section we discuss the design of our CoAP over ICN architecture, alongside the benefits our architecture can offer to CoAP.


\subsection{Design}

\begin{figure}
\centering
\includegraphics[width=0.9\linewidth]{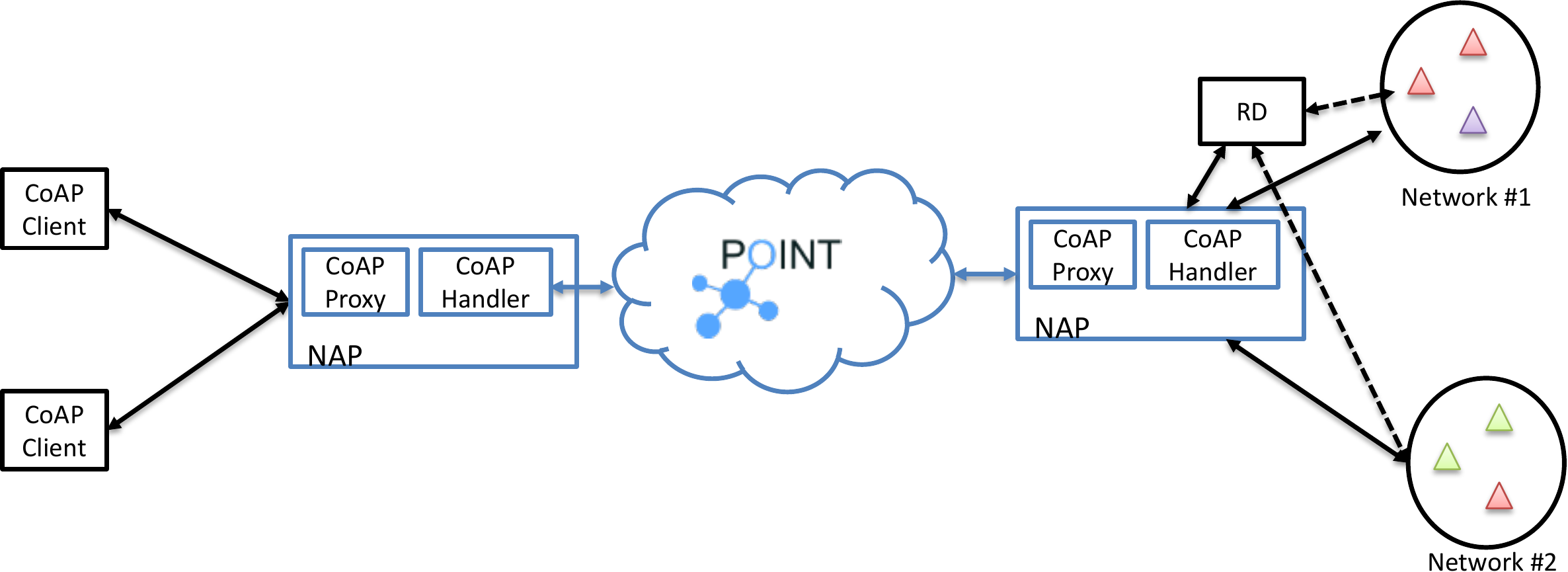}
\caption {Our reference architecture. On the right part there are Things offering resources. Each resource is specified by a color. On the left part there are CoAP clients.}
\label{fig:arch}
\end{figure}

Figure~\ref{fig:arch} illustrates the design of our CoAP over ICN architecture. In the middle of the figure there is the POINT network that interconnects NAPs. In the core of the
POINT network the PSI architecture has been deployed.In the right part of the figure there are networks of \textit{Things}. 
Each Thing acts as a CoAP server offering a resource; the same (type of) resource can be offered by many Things located in different networks 
(e.g., there can be many sensors deployed in various parts of a city offering temperature measurements). Each network of Things is connected to the POINT network through 
a CoAP proxy implemented in a NAP. 
A CoAP Resource Directory (RD) hosts the descriptions of resources provided by the CoAP servers; 
CoAP servers have to register resources to the RD with standardized resource descriptions. 
NAPs learn available resource names by communicating with RDs and \emph{subscribe} to the appropriate ICN identifier. This identifier is derived using the \emph{host-uri} of the CoAP resource. 
In the left part of the figure there are CoAP clients. A CoAP client is also connected to the POINT network though a CoAP proxy implemented in a NAP.

The fundamental component of our CoAP over ICN architecture is a \textit{CoAP handler} which is part of the NAP. A CoAP handler receives CoAP requests from CoAP clients (over IP), translates them into ICN messages, and advertises them the to the ICN network. The identifier of these messages includes the \emph{host-uri} of the CoAP resource, therefore they trigger the
ICN rendezvous process, which eventually leads to the forwarding of these  messages to the appropriate NAP(s). A NAP that receives an ICN message restores the original CoAP request and forwards it to an appropriate CoAP server. The CoAP server generates a response which is forwarded through the ICN network to the CoAP clients following the reverse process.

\begin{figure}[h]
\centering
\includegraphics[width=.95\linewidth]{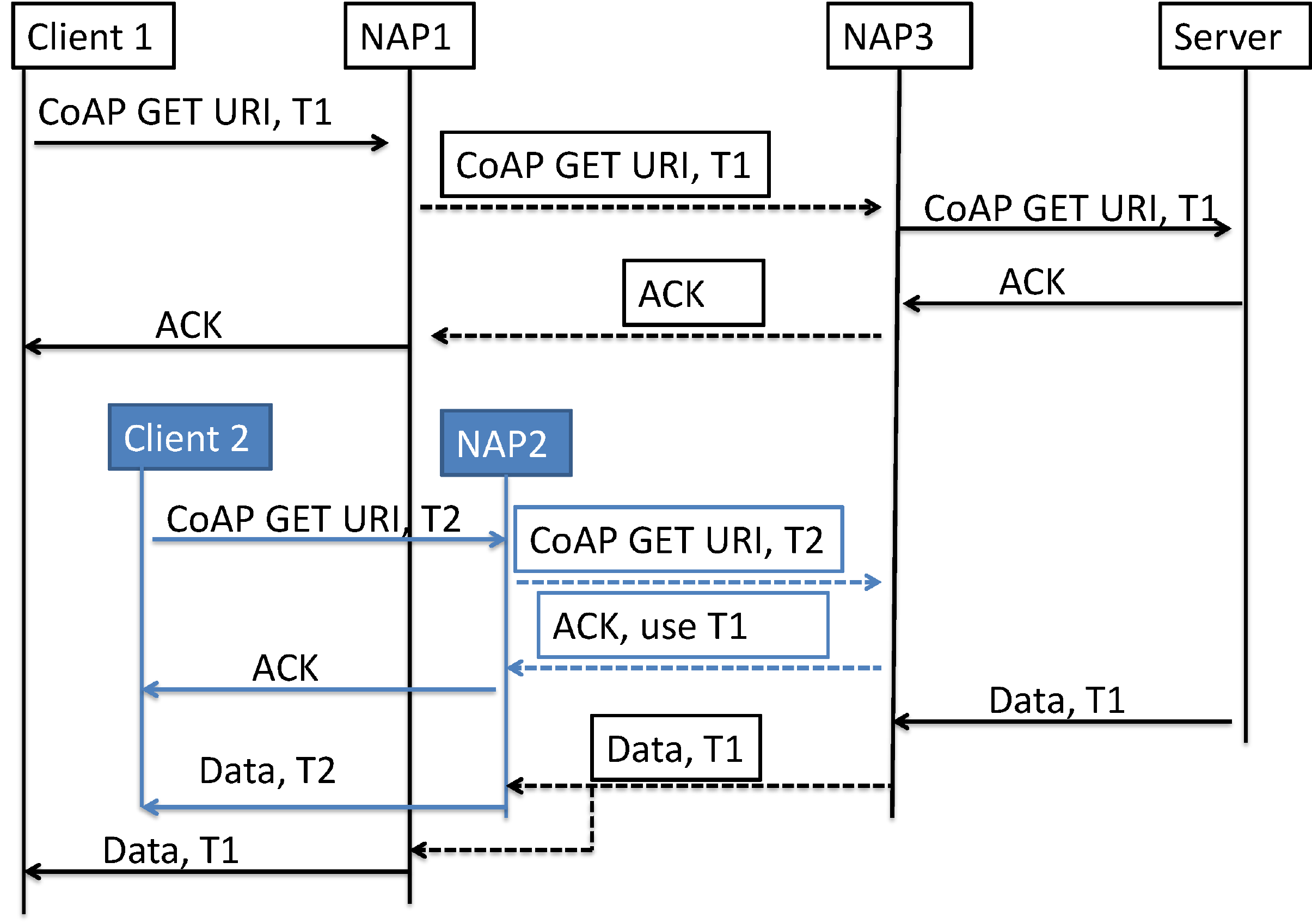}
\caption {Coincidental multicast. Dotted lines are ICN messages transferred inside the ICN network}
\label{fig:multi}
\end{figure}

\subsection{Benefits of ICN to CoAP}
We now present some typical CoAP communication scenarios and we discuss the benefits of ICN underlay to CoAP.

\subsubsection{Coincidental multicast}
CoAP client may issue a request for a resource that is not yet available. In this case, the CoAP server will acknowledge the request and when the resource becomes available, it will send the appropriate response back to the client. The response should contain the same token as the request, so as the client to be able to correlate them. If multiple clients request a resource not yet available, the CoAP server should maintain state and could  \emph{simultaneously} respond to all these requests when the resource becomes available. A similar behavior occurs when the CoAP \emph{observe} extension is used. Many CoAP clients may register their interest in a resource; when the state of a resource changes, the CoAP server could notify simultaneously all CoAP clients. In typical IP networks this communication pattern would result in multiple unicast transmissions from the CoAP server to the CoAP clients. In contrast, in POINT the impact to the network of this type of bursty traffic can be reduced by employing multicast. In order to achieve this, the POINT network instructs NAPs to use the same token for all these identical CoAP requests. NAPs are then responsible for modifying the token to the CoAP response in order to match this to the original CoAP request sent by the CoAP endpoint. 

Figure~\ref{fig:multi} shows an example of coincidental multicast. In this case a CoAP client has requested a resource, including Token``T1'' in the request. NAP1 translates the CoAP request into an ICN message and forwards it to the network. NAP3 receives the ICN message, translates it into a CoAP request and forwards it to the appropriate CoAP server. We assume here that the resource is not yet available, therefore the CoAP server simply ACKs the request. The acknowledgment is eventually received by the CoAP client. 
After some time another CoAP client requests the same resource through NAP2. This client includes in his request another token, namely ``T2''. Similarly to the previous message, NAP2 translates the CoAP request into an ICN message and sends it to the network. This time, when NAP3 receives this message, it will not contact the CoAP server, instead it will inform NAP2 to expect a response that will have in its payload the token ``T1''. When the resource becomes ready, the CoAP server sends a single unicast CoAP message to NAP3. NAP3 translates it into an ICN message and sends it using multicast to NAP1 and NAP2.\footnote{How the multicast delivery tree is constructed is out of the scope of this paper. Interested readers are referred to~\cite{Xyl2012} for how this can be achieved.}. Finally, NAP1 and NAP2 will transform the ICN messages into the appropriate CoAP responses, including the correct tokens.

\subsubsection{One-to-many requests}
CoAP group communication allows a CoAP client to send a request to a group of CoAP servers. A CoAP group name may have application specific semantics (e.g., sensors.west.building6 which translates to all sensors located in the west wing of building 6). In an IP network this functionality can be implemented by creating an IP multicast group per CoAP group and by modifying DNS to translate from CoAP group names to IP multicast addresses. Therefore, the steps required to send a CoAP group request are: (i) CoAP servers join a number of IP multicast groups, (ii) CoAP group name to IP multicast address records are added to the appropriate DNS servers, (iii), a CoAP client (or the forward proxy) that wants to send a CoAP request performs a DNS resolution and learns the IP multicast destination address for the request, (iv) a CoAP server that receives a request from a multicast group, examines if it offers the resource included in the request and decides whether or not to respond (based on the application context).

It can be observed that group communication adds some overhead to CoAP endpoints: they have to support IP multicast, they have to maintain extra state, and they may receive redundant requests. On the contrary, relying on ICN, the POINT network can translate automatically a CoAP \emph{request} to endpoint locations. Therefore, there is no need for DNS resolution and IP multicast support because ICN operations consider the complete request, not only the group name, and there are no redundant messages.  
The POINT network could also provide anycast requests, i.e., forward a request to the most ``suitable'' endpoint. The decision about which endpoint is suitable depends on various aspects e.g.,  CoAP client QoS requirements, chances for coincidental multicast, load balancing, etc.

\section{Evaluation Platform}
\label{sec:plat}

\subsection{Evaluation criteria}
The preliminary evaluation includes how much benefits POINT platform offers to CoAP in terms of latency, state management, overhead in the network as compared to vanilla TCP/IP. The next phase of this work is to add a cross protocol proxy to the CoAP handler that will provide a mapping between CoAP and HTTP. The key objective of our evaluation is to evaluate the proposed architecture based on the hypothesis that running IP-over-ICN results in a better networking experience for the major stakeholders, compared to the traditional TCP/IP networking. 

\subsection{Prelimenary evaluation}
Early indications show that the POINT architecture offers advantages to CoAP application developers, as well as, to operators. When POINT is used, a CoAP application does not have to deal with IP multicast. Moreover, no modifications to DNS are required. The POINT network transfers state overhead from the CoAP (constrained) endpoints to the network. In particular, in case of requests to not yet available resources, as well as, when the CoAP observe extension is used, the CoAP server receives a single request, since all other requests are suppressed by the NAPs. As far as the operator is concerned, CoAP and CoAP observe create many chances for multicast communication; the POINT network then takes advantage of this and uses multicast to handle bursts of traffic.

\subsection{Experimental evaluation environment}
To evaluate the performance of the CoAP over ICN solution, we have constructed an IoT testbed which is connected to the existing POINT testbed through a NAP. The POINT testbed connects all partners of the POINT project throughout Europe using OpenVPN. The POINT overlay testbed is based on Blackadder ICN platform, which has already demonstrated ICN performance at data rates up to 10 Gb/s \cite{riihijarvi2012final}. The IoT testbed consists of sensors attached to single-board computers, such as Nucleo boards with ELL-i Ethernet NICs and the RIOT operating system~\cite{riot}, as well as  Raspberry Pis. These single-board computers run a CoAP server. In addition to the actual sensors, we are planning to use a sensor traffic generator (STG).\footnote{Source code available at https://github.com/vr000m/SensorTrafficGenerator} The STG is capable of emulating the traffic of various types of sensors, e.g., light and temperature sensors, GPS receivers, and cameras.

\section{Conclusions and Future work}
\label{sec:conc}
CoAP functionality and principles are very close to those of ICN: asynchronous communication, persistent interests, and group communication, all are intriguing communication paradigms that are found in many ICN architectures. Therefore, using ICN to transport CoAP traffic seems a natural choice. The work presented in this paper is still in early stages. Yet, even from these first steps some advantages of ICN for CoAP and to CoAP providers are apparent. Future releases of the POINT platform will incorporate the CoAP functionality described in this paper. Moreover, it is in our immediate plans to seamlessly support CoAP over DTLS (see Section 9.1 RFC 7252). 
%

Some research issues may extend beyond the POINT project: what benefits would materialize if all the local networks the IoT devices attach to are ICN, or use ICN underlays? This would open up the possibility for dynamic roaming and automatic meshing to facilitate better failure resilience for critical traffic. Can the endpoints still use CoAP over IP, or do they need to be ICN aware?

For more practical matters, using strong cryptography is a must for devices controlled over the Internet. The key distribution is always a problem and provisioning large numbers of devices will make that even harder to correctly set up Things. Therefore the need for human intervention should be minimized and bootstrapping trust by leveraging ICN specific security mechanisms (as for example in~\cite{Pol2015}) is an interesting future work direction.

\section*{Acknowledgments}
Many of the ideas presented in this paper stem from discussions among POINT consortium partners. 
The work presented in this paper was supported by the EU funded H2020 ICT project POINT, under contract 643990.
\bibliographystyle{IEEEtran}

\bibliography{IEEEabrv,ntms}

\end{document}